
\magnification=\magstep1
\hsize=5.9truein
\vsize=8.375truein
\hoffset=.45in
\parindent=25pt
\nopagenumbers
\raggedbottom
\hangafter=1
\def\makeheadline{\vbox to 0pt{\vskip-40pt
   \line{\vbox to 8.5pt{}\the\headline}\vss}\nointerlineskip}
\def\approxlt{\kern 0.35em\raise 0.45ex\hbox{$<$}\kern-0.66em\lower0.5ex
   \hbox{$\scriptstyle\sim$}\kern0.35em}
\def\approxgt{\kern 0.35em\raise 0.45ex\hbox{$>$}\kern-0.75em\lower0.5ex
   \hbox{$\scriptstyle\sim$}\kern0.35em}
\vglue1.5truein
\centerline{\bf QCD Instantons and 2D Surfaces  }
\vskip22pt
\centerline{V.G.J. Rodgers}
\centerline{ Department of Physics and Astronomy}
\centerline{ The University of Iowa}
\centerline{ Iowa City, Iowa~~52242--1479}
\centerline{ Jan 1992 }

\vglue0.5truein
\centerline{Dedicated to the memory of}
\centerline{ Prof. Dwight Nicholson,
Prof. Bob Smith, Prof. Chris Goertz,}
\centerline{ Dr. Anne Cleary and
Dr. Linhua Shan,}
\centerline{ and for the recovery of Ms. Miya Sonya Sioson.}
\vglue0.5truein
\baselineskip=12pt
\centerline{\bf ABSTRACT}
\vskip22pt
Some time ago, Atiyah showed that there exists a natural identification
between the k-instantons of a Yang-Mills theory with gauge group
$G$ and the holomorphic
maps from $CP_1$ to $\Omega G$.  Since then, Nair and Mazur,
have associated the $\Theta $ vacua structure in QCD with self-intersecting
Riemann surfaces immersed in four dimensions.  From here they concluded that
these 2D surfaces correspond to the non-perturbative phase of QCD and
carry the topological information of the $\Theta$ vacua.  In this paper
we would like to elaborate on this point by making use of Atiyah's
identification.   We will argue that an effective description of QCD may
be more like a $WZW$ model coupled to the induced metric of an immersion of
a 2-D Riemann surface in $R^4$.  We make some further comments on the
relationship between the coadjoint orbits of the Kac-Moody group on $G$
and instantons with axial symmetry and monopole charge.
\vfill
\eject
\headline={\tenrm\hfil\folio}
\baselineskip=20pt
\pageno=1
One of the outstanding questions that still haunts theoretical
physics today is that of the nature of the non-perturbative phase of  QCD.
Probes such as lattice calculations, $1/N_c$ expansions and topology,
are beginning to reveal  some of the salient features of the
rich vacuum structure of QCD.  In particular it is believed that a
string theory may provide an effective description of  QCD in this
non-perturbative phase.   Furthermore, it has been suggested that
a string theory in four dimensions  could enjoy a $\Theta$-like term
[1,2], and that this $\Theta$ term can be associated with the $\Theta$
vacua of  QCD [3].  This term would correspond to the self-intersections
of the image of a Riemann surface immersed in $R^4$.  Thus one is able to
capture an important feature of non-perturbative QCD in a string
theory.  In this note we would like to extend this list of qualitative
features to include a correspondence between the moduli space of the maps
associated with a 2D theory and the moduli space of the instanton sector
of  QCD.  Indeed it has been known for some time [4] that not only is there
a general correspondence between 2D structures and 4D Yang-Mills instantons,
but that there also exists a diffeomorphism between the moduli spaces of
certain maps from $CP_1$
into the loop group of $G$, and the moduli space of instantons on $R^4$.
More specifically, Atiyah [4] has shown that the moduli space of 4D
$k$-instantons is diffeomorphic to the moduli space of degree k, holomorphic
maps from $CP_1$ into the loop group of $G$, viz $\Omega G$, where $G$ is the
gauge group of the Yang-Mills theory.  Thus one has a natural identification
between 4D instantons and 2D surfaces.

We will exploit this identity to write down an action that may correspond to
the instanton sector of  QCD.  First let us recall the overall features
of the work of Atiyah.  The main emphasis is on holomorphic maps, $f$, which
carry $CP_1$ into $\Omega G$, where $\Omega G$ in the loop group of the
gauge group $G$ of the Yang-Mills theory in question.  The space
$\Omega G$ consists of all maps from $S^1$ into $G$, $g: S^1\rightarrow G$,
such
that $g(\theta = 0) = 1$.  Thus the only constant map is the
identity map.  We will, of course,
be primarily interested in the case when $G$ is $SU(3)$ although the work can
be extended to more general cases.  Two important features
of $\Omega G$ are that it is a K\"ahler manifold, and that it is
infinitely dimensional.  The relationship between the maps $f: CP_1 \rightarrow
\Omega G$ and the Yang-Mills instantons arise from the fact that

a) By twistor methods, 4D instantons are described by holomorphic maps,
 $f: CP_1 \rightarrow \Omega G$ and

b) by using the results of Donaldson [5], one can show that all such
holomorphic maps arise from these twistor methods.  (Although the
argument is proved only for classical groups, this is sufficient
for our purposes.)

{}From the two statements above , Atiyah is able to prove the following
theorem;

{\bf Theorem}[Atiyah]:
The moduli space of Yang-Mills k-instantons (k an integer) over $R^4$
with classical group $G$, modulo based gauge transformations and the
moduli space of {\it all} based holomorphic maps from $CP_1 \rightarrow
\Omega G$ of degree k, are diffeomorphic to each other.  We will denote
by ${\cal M}_{4D}$ and ${\cal M}_{2D}$ for the respective moduli spaces.

We note that  the Yang-Mills moduli space mentioned in the above
theorem is that which includes moduli associated with constant
gauge transformations.  To clarify this point, consider $G=SU(2)$.
By acting on known solutions of the equations of motion with the
symmetries of the Yang-Mills action one can find new solutions that
may not be gauge related.   Indeed by using the
full conformal group (Poincar\`e transformations, and the conformal
transformations) one can show that there are four parameters coming from
the translations, one coming from the scale transformations, and three coming
from the action of the constant gauge transformations.  Thus there is an
$8k$ dimensional moduli space for the k-instanton solutions.  If we ignore
the constant gauge transformations, this is tantamount to using all the
gauge transformations (including the constant ones) in the equivalence
relations.  This effectively reduces the dimension of the moduli space to
$8k-3$ dimensions [6].  This moduli space is denoted by ${\cal M}_{4D}/G$
and in general may not be a manifold due to a finite set of points that
are related to the reducible connections.

Now let us focus on $\Omega G$.  It is well known that as a manifold
$\Omega G$ is K\"ahler [7].  Furthermore, by using methods of Kirillov [8], one
can show that it has a natural symplectic structure associated with
a particular coadjoint orbit, namely the orbit associated with a
pure centrally extended covector.  In Ref. [9,10] we used this symplectic
geometry to
construct an action that on $CP_1$ is directly related to the
natural K\"ahler metric on $\Omega G$.  By using the Kac-Moody algebra,
which is the centrally extended algebra of $\Omega G$, the action
necessarily contains order $\hbar$ contributions.  These order $\hbar$
contributions will be necessary for a realistic effective action of  QCD.

The maps that we seek for the string action for QCD will be maps from
$CP_1 \rightarrow \Omega G$. By using a general procedure [11],
 one is able to consider a two-dimensional
manifold $ m$ where each point on this manifold corresponds to that
element in $\Omega G$ which is used to transport a coadjoint vector, {\bf b},
to a new element $ {\bf b_g}$ on the coadjoint orbit.  The fields, $g(\sigma,
\lambda,\tau)$, provide a map from  $m$ to $\Omega G$.
By a suitable choice of boundary conditions  we
may take $m$ to be $CP_1$.  Now it is the coadjoint orbit  that is
endowed with the symplectic structure.  One can show explicitly that the
action associated with the symplectic structure of $\Omega G$ is just a
WZW model corresponding to the orbit
of the covector ${\bf b_0}=(0,\mu)$.  In general, there will be an extra
term added to the WZW action for those covectors where $ {\bf b}=(B\not=
0,\mu)$
and which cannot be transported to ${\bf b_0}=(0,\mu)$ by loop group action.
The
existence of these orbits constitutes the difference between $\Omega G$,
which is the space of based loop on G, and ${\cal G}$, which is the space
of free loops on G.  Indeed the constant loops which correspond to
group elements in G which are not the identity live in ${\cal G}$.
By choosing the coadjoint orbit with ${\bf b_0}=(0,\mu)$ we are
looking at the symplectic structure on ${\cal G}$ which is invariant under all
constant group transformations.  Thus $\Omega G = {\cal G}/G$ acquires this
symplectic structure.  Let $\omega_b$ be the symplectic two-form associated
with the orbit of a covector {\bf b}.  Then the action we seek is simply
$$S = \int_{CP_1} \omega_b.$$
This action for ${\bf b_0}=(0,\mu)$ will be defined over maps from
$CP_1 \rightarrow \Omega G$.  In general for a generic covector we may
explicitly write,
$$
\eqalign{ S_0= {n \over 2 \pi} & \int Tr \left\{ g
Bg^{-1} \left[ {{\partial g}\over {\partial
\lambda}}g^{-1}, {{\partial g}\over{\partial \tau}}g^{-1}\right] \right\}d
\sigma d \tau d \lambda \cr
\noalign{\vskip5pt}
 + {{n}\over{2 \pi}} & \int \mu Tr \left\{ {{\partial g}\over{\partial
\sigma}}g^{-1}
\left[ {{\partial g}\over{\partial \lambda}}g^{-1}, {{\partial g}\over{\partial
\tau}}g^{-1} \right] \right\}d \sigma d \tau d \lambda  \cr
\noalign{\vskip5pt}
 + {{n}\over{2 \pi}} & \int \mu Tr \left\{ {{\partial g}\over{\partial
\lambda}}g^{-1} {{\partial}\over{\partial \sigma}} \left( {{\partial
g}\over{\partial
\tau}}g^{-1} \right) \right\}d \sigma d \tau d \lambda}
$$
where the integration is  over a 3-manifold with $ CP_1 \times S^1 $ topology.
The $S^1$ factor is coming from the loop integration.    Integrating by parts
and ignoring total $\tau$ derivatives we have

$$
\eqalign{  S_0= & {{-n}\over{2 \pi}} \int_\Sigma Tr \left( Bg^{-1} {{\partial
g}\over{\partial \tau}} \right)d \sigma
d \tau  \cr
\noalign{\vskip5pt}
    & - {{n \mu}\over{4 \pi}} \int_\Sigma Tr g^{-1} {{\partial g}\over{\partial
\sigma}}g^{-1}
{{\partial g}\over{\partial \tau}}d \sigma d \tau  \cr
\noalign{\vskip5pt}
   &  + {{n \mu}\over{4 \pi}} \int_{CP_1\times S^1} Tr g^{-1} {{\partial
g}\over{\partial \sigma}} \left[ g^{-1}
{{\partial g}\over{\partial \lambda}}, g^{-1} {{\partial g}\over{\partial
\tau}} \right]d
\sigma d \tau d \lambda.}
$$
Here $\Sigma$ is a two-dimensional surface that remains after the integration
by parts.

Now just as stated the isotropies of the field {\bf b} will be used to
determine which coadjoint orbit we choose.  The isotropy group of {\bf b} will
arise from those group elements that leave {\bf b} invariant.  For
${\bf b}= ( B, \mu)$, this is just the statement that there exists elements,
$h(\sigma) \in {\cal G}$ such that
$$
h(\sigma): (B(\sigma),\mu) \equiv \biggl(h(\sigma) B h^{-1}(\sigma) + \mu
{\partial
h \over \partial \sigma} h^{-1}, \mu\biggr) = (B,\mu). $$
Thus the isotropy group of {\bf b} specifies the field space of the
action.  The orbit of {\bf b} is thus identified with all those elements
of ${\cal G} \simeq G \times \Omega G$ modulo $H$, or in other words
$G/H\times \Omega G$.  Generally this is because only constant elements
in ${\cal G}$ will provide isotropy for {\bf b} when $\mu\not= 0$.  The purely
central extension covector,
${\bf b_0}=(0,\mu)$, defines the coadjoint orbit that is identified
with $\Omega G$.

This established the first phase of our search for an effective QCD action.
We have successfully identified a 2D action (in the WZW sense) with that
of the instanton sector of Yang-Mills.  However, as it stands the action
$S_0$ is not enough for a 4D effective action for QCD.  For one thing
we need a 4D description and for another we would like to represent the
$\Theta$ vacua.

We can write an action with these attributes and also incorporate the
action $S_0$ if we consider the immersion of 2D surfaces into a 4D
manifold that corresponds to the image of the two surface $\Sigma$
and the image of the (WZW) manifold with $CP_1 \times S^1$ topology
for the WZ term.  One can consider the induced metric and
antisymmetric tensors arising from the immersions as the background
geometric objects which are coupled to the WZW model with $SU(3)$
Kac-Moody symmetry.  Since we are allowing immersions, as opposed to
embeddings, self-intersecting surfaces corresponding to the image
of $\Sigma$ can be included.  These self-intersecting surfaces will
provide the means by which the analogue of the $\Theta$ vacua may arise [3].

By writing the WZW model in Lorentz invariant form and expressing the
coadjoint vector {\bf b} as a world surface vector (recall that
the symplectic geometry fixes a gauge), ${\bf b}=(B_a,\mu)$, we may
write our proposed action as,
$$
\eqalign{
I= & \displaystyle \int d^2 x Tr B_{a}  g^{-1}\partial _{b} g ( \sqrt{H} H ^{a
b}+E^{a b}) \cr
\noalign{\vskip5pt}
+ & {{1}\over{24 \pi}} \int d^2 x Tr \partial _{a} g \partial_{b}
g^{-1} \sqrt{H} H^{a b} \cr
\noalign{\vskip5pt}
+ &  \displaystyle {{1}\over{24 \pi}} \int d^3 y E^{ijk} Tr (g^{-1} \partial_ig
g^{-1} \partial_jgg^{-1} \partial_kg) \cr
\noalign{\vskip5pt}
+ & \displaystyle \Theta {-1\over 16\pi}\int d^2 x \sqrt{H} H^{a b}
\partial_a \tau^{\mu \nu} \partial_b \tau^{\lambda \rho}
\epsilon^{\mu\nu\lambda\rho}}.
$$
In the above, the induced tensors are defined through the relations,
$$
\eqalign{
t^{\mu}_a &\equiv \partial_a X^{\mu}( \sigma,\tau ) \cr
H_{ab} &\equiv t^{\mu}_a t^{\nu}_b G_{\mu\nu} \cr
E_{ab} &\equiv t^{\mu}_a t^{\nu}_b N^{\lambda}_A N^{\rho}_B \epsilon^{AB}
\epsilon_{\mu\nu\lambda\rho } \cr
E_{ijk} &\equiv t^{\mu}_i(\sigma,\tau,\lambda) t^{\nu}_j(\sigma,\tau,\lambda)
t^{\lambda}_k(\sigma,\tau,\lambda) \zeta^{\rho} \epsilon^{\mu\nu\lambda\rho}
},$$
where the latin indices, $a,b\in (1,2)$, $A,B\in (1,2)$ and $i,j,k\in (1,2,3)$,
while the
Greek indices are 4D space-time indices.  The $ t^{\mu}_a $'s span the tangent
space bundle on the image of $\Sigma$, and the $N^{\rho}_A$'s span the $SO(2)$
normal bundle.  For the WZ term we use the tangent vectors,
 $t^{\mu}_i, i=1,2,3$ for the image of the $ CP_1 \times S^1 $ and
$\zeta^{\rho}
$ as the outward drawn normal to the three surfaces.  We denote by $H$ the
determinant of $H_{ab}$.  For the sake of generality, we have considered a
four manifold with metric $G_{\mu\nu}$ and orientation tensor
$\epsilon^{\mu\nu\lambda\rho}$.  The first term of the action specifies the
space of fields where the action is defined.  We are interested in the
field space $\Omega G$.  Thus we may set $B_a = 0$ and $\mu={1\over 2}$.
Setting $\mu={1\over 2}$ and $n=1$ from the WZW model is sufficient to
guarantee a consistent quantum theory [12]. The second and third terms of
the action are the usual WZW model coupled to the image of the
Riemann surface $\Sigma$ which is specified by $X^{\mu}(\sigma,\tau)$.
The 3D WZ term is defined on an image of $CP_1 \times S^1$ which is specified
by the extended immersion vector $X^{\mu}(\sigma,\tau,\lambda)$.  This
WZ term is coupled to the induced antisymmetric rank three tensor, $E^{ijk}$,
on the image of $CP_1 \times S^1$.  The last term is the self-intersecting
term for the image of the 2D surface, $\Sigma$.  It is this term that will
capture the $\Theta$ vacua structure.  The self-intersecting surfaces
will act as the n-vacua and the instantons will arise as solutions to
the Euler-Lagrange equations of the WZW model.  For each of the
self-intersecting n sectors, there will be the lowest lying state
that will serve as the vacuum.  From what we know of ordinary strings
in 4D, these lowest lying states may correspond to torus knots [13].
We will examine this claim in a later publication.

Let us remark that upon quantization a Liouville mode should emerge.  This
will break the scale invariance of the theory.  We conjecture that this
introduction of a scale is necessary and that it should be consistent
with the scale for the non-perturbative phase for QCD.  Indeed it would
be interesting if $\Lambda_{\rm QCD}$ were related to the vacuum expectation
value of the Liouville mode from a string theory.  Since the scale arises
from the anomaly associated with the conformal invariance, after gauge
fixing to the light cone gauge the new
action will have an extra contribution of [14]
$$
I~'_{0}= {1 \over 21 \pi } \int d \sigma d \tau \left[
{{\partial^2_{\sigma} s}\over{(\partial_{\sigma}s)^2}} \partial
_{\tau} \partial_{\sigma} s - {{(\partial^2_{\sigma}s)^2
(\partial_{\tau} s)}\over{(\partial_{\sigma} s)^3}} \right].
$$
In the above $s(\sigma,\tau)$ is the Liouville mode.  The coefficient
of the above comes from using $n=1, c_v=12$, and Dim $G = 8$ in
$ {2n~ {\rm Dim} G\over 2n + c_v} $  [14].

We have proposed the action $I$ as a possible effective action for QCD in its
non-perturbative phase.  The action is motivated by an observation of Atiyah
that shows the moduli spaces for instantons and certain 2D maps
enjoy a diffeomorphism between them.  Physically, we expect the
vacuum structure of QCD, with regard to the topological structure of the
$SU(3)$ gauge group, to remain unchanged during a phase transition.
In other words, the existence of instantons and the $\Theta$ vacua are
generic properties of the $SU(3)$ bundle over $R^4$.  It is plausible
to assume that as QCD undergoes strong coupling into a string-like
phase, that the vacuum topological structure remains unaltered.  If
this is true then one must consider a string theory that, at the
very least, preserves the moduli structure of instantons and which
has features exhibiting the $\Theta$ vacua.  We have constructed such
a string theory here.  The partition function will require integration
over the fields $g(\sigma, \lambda, \tau)$ and all possible immersions
that admit a differential structure.  The partition function will
be
$$
Z=\int {\cal D}X^{\mu} {\cal D}g ~e^{-I(X^{\mu}, g)}
$$
where we have ignored gauge fixing terms and the conformal anomaly.
We will address the quantization of this model in a later publication.

As a final remark, one may ask what is the analogue for the
${\bf b}=(B\not=0,\mu)$ orbits in terms of instantons?  As we
stated earlier, different covectors may correspond to different orbits
where the orbits may be identified with
$ {\cal G}/H \simeq G/H \times \Omega G$.
Let $B$ be the generator of a homomorphism from the circle $S^1$ into
G. For example when $G=SU(2)$ we may have
$$B=\pmatrix{n&0\cr
0&-n\cr} $$
 which
generates the homomorphism
$$\alpha(\sigma)=\pmatrix{e^{i n \sigma}&0\cr 0&e^{-i n \sigma}\cr}.$$
Thus $B$ has a $U(1)$ isotropy group.  The coadjoint action of
$g(\sigma,\lambda,\tau)$ (which includes constant loop transformations
which are not the identity) on {\bf b} will provide maps from
$CP_1 \rightarrow G/H \times \Omega G$.  Now in a subsequent theorem, Atiyah
[4] has
shown that there exists an isomorphism between $S^1$ invariant $k$
instantons which are associated with a homeomorphism $\alpha$ and the
parameter space of based holomorphic maps from $CP_1 \rightarrow G/H$
of degree $k$, where $H$ is the centralizer of $\alpha$.  Note that
one specifies $\alpha$ in the above theorem since there may be many ways to
embed $H$ in $G$.  Thus inequivalent instantons may have the same isotropy
group.  With this theorem, we are able to identify the orbit of a
covector {\bf b} with isotropy group $H$ as an axial symmetric instanton.
Furthermore, Atiyah has shown that such $S^1$ invariant $k$-instantons are
monopoles on the 3-space $H^3$.  Our interpretation is then that the other
orbits  are related to axial symmetric instantons
which can be identified with colored monopoles with instanton charge.
These monopoles will in general break the global $SU(3)$ color symmetry
and this phenomenon is quite reminiscent of the work found in [15] on
global color breaking due to monopoles.  In general there will be several
inequivalent orbits with the same coset space $G/H$ describing the field
space due to the number of ways one can embed the subgroup $H$ into $G$.
Thus we have captured another topological aspect of QCD in this framework,
namely magnetic monopoles.  Again this is consistent with our philosophy
that topological properties associated with the gauge group will have
string analogues.  For an interesting view of why all instantons are
monopoles see Ref. [16].

Just as we expect a Liouville mode in the zero monopole sector we should
expect it to appear in the non-zero sector as well.  Indeed [14] this
is true and one finds that after gauge fixing
in the light cone
gauge there is an extra contribution to the action due to the conformal
anomaly in the monopole sector,  i.e.
{\bf b}$=(B, {1\over 2}$) is such that $B\not= 0$ or any pure gauge
configuration,  that we may write this contribution as
$$\eqalign{
I~'_{B}=   {1 \over 21 \pi } & \int d \sigma d \tau \left[
{{\partial^2_{\sigma} s}\over{(\partial_{\sigma}s)^2}} \partial
_{\tau} \partial_{\sigma} s - {{(\partial^2_{\sigma}s)^2
(\partial_{\tau} s)}\over{(\partial_{\sigma} s)^3}} \right] \cr
\noalign{\vskip5pt}
+ & \int B(\sigma) \left( {\partial_\lambda s\over
\partial_\sigma s} \partial_\sigma \left[ g^{-1}
\partial_\tau g\right] -  {\partial_\tau s\over \partial_\sigma s}
\partial_\sigma \left[ g^{-1} \partial_\lambda g \right] \right) d\sigma~
d\lambda~ d\tau \cr
\noalign{\vskip5pt}
+& \int B(\sigma) \left[ g^{-1} \partial_\lambda g, g^{-1} \partial_\tau g
\right]
 d\sigma~ d\lambda~ d\tau. \cr
}
$$
In the above $g=g(\sigma,\lambda,\tau)$.  The last two terms resemble
a $BF$ system where the derivative operators are modified by the gauge
fixed 3D extended metric so that $g^{-1} \partial g$ is no longer
pure gauge.
We will try to clarify the role of these monopoles as $BF$ and Chern-Simons
theories in Ref. [14].

The work above can of course be put on a more general footing than
QCD.  One can study the instanton sector of other theories using
these ideas.
Finally we would like to note that P.A. Griffin has also considered the
role of WZW models in explaining the non-perturbative phase of
QCD through lattices techniques [17].

Acknowledgements:  I would like to thank A.P. Balachandran, J. Gates,
R. Lano, V.P. Nair, P. Majundar and Y. Meurice for discussion.
I would also like to thank The Institute for Mathematical Sciences,
T.S. Krishnaswamy, T.S. Narain, T.S. Raman, T.S. Ganapathy,
and T.S. Srinivasan
for hospitality while in Madras, India.  This work was supported in
part by NSF Grant PHY-9103914.

\vfill\eject
\nopagenumbers
\baselineskip=13pt
\centerline{\bf REFERENCES}
\settabs 1 \columns

\+ [1]  A.P. Balachandran, F. Lizzi, and G. Sparano, Nucl. Phys.B263 ( 1986)
608 \cr
\+ [2]  A. Polyakov, Nucl. Phys B268 (1986) 406 \cr
\+ [3]  P.O. Mazur and V.P. Nair, Nucl. Phys.B284 (1986) 146 \cr
\+ [4]  M. Atiyah, Comm. Math. Phys. 93 (1984) 437 \cr
\+ [5]  S.K. Donaldson Comm. Math. Phys. 93 ( 1984) 453 \cr
\+ [6]  For a review see, S. Coleman, ``The Uses of Instantons'', Erice
Lectures 1977   \cr
\+ [7]  A. Pressley, and G. Segal, {\it Loop Groups}, Oxford 1988 \cr
\+ [8]  A.A. Kirillov, {\it Elements of the Theory of Representations}
 ( Springer-Verlag, 1975) \cr
\+ [9]  B. Rai and V.G.J. Rodgers, Nucl Phys. B341 (1990) 119 \cr
\+ [10]  G.W. Delius, P. van Nieuwenhuizen, and V.G.J. Rodgers,\cr
\+ ~~~~~Inter. Jour. of Mod. Physics A5 (1990), 3943 \cr
\+ [11] F. Zaccoria, E.C.G. Sudarshan, J.S. Nilsson, N. Mukunda, G. Marmo,\cr
\+ ~~~~~and A.P. Balachandran,  Phys. Rev D27 (1983) 2327;\cr
\+ ~~~~~A.P. Balachandran, G. Marmo, B.S. Skagerstam, and A. Stern,\cr
\+ ~~~~~{\it Gauge Symmetries and Fibre Bundles; Applications to Particle
Dynamics},\cr
\+ ~~~~~Springer-Verlag Berlin (1983)\cr
\+ [12] A.P. Balachandran, G. Marmo, A. Stern, Nucl. Phys. B162 (1980) 385 \cr
\+ ~~~~~A.P. Balachandran, G. Marmo, B.S. Skagerstam, and A. Stern,\cr
\+ ~~~~~Nucl. Phys. B164 (1980) 427 \cr
\+ ~~~~~J.L. Friedman and R. Sorkin, Comm. Math. Phys. 73 (1980) 161 \cr
\+ ~~~~~E. Witten, Nucl. Phys. B233 (1983) 422 \cr
\+ [13] G.D. Robertson, Phys. Lett. B226 (1989) 244 \cr
\+ [14] V.G.J. Rodgers, Manuscript in preparation \cr
\+ [15] A.P. Balachandran, G. Marmo, N. Mukunda, J.S. Nilsson,\cr
\+ ~~~~~E.C.G. Sudarshan, F. Zaccaria \cr
\+ ~~~~~Phys.Rev.Lett.50 (1983) 1553, Phys.Rev.D29 (1984) 2919, Phys.Rev.D29
(1984) 2936 \cr
\+ [16] H. Garland and M.K. Murray, Comm. Math. Phys. 121 (1989) 85 \cr
\+ [17] Paul A. Griffin, ``Solving 3+1 QCD on the Transverse Lattice \cr
\+ ~~~~~Using 1+1 Conformal Field Theory'', FermiLab Preprint 91/197-T \cr
\end